\documentclass[a4paper,11pt,twocolumn]{article}
\usepackage{cite}
\usepackage[T1]{fontenc}
\usepackage{graphicx}
\usepackage{authblk}
\usepackage{amsmath}
\usepackage[margin=0.8in]{geometry}

\usepackage{array}
\usepackage{caption}

\newcolumntype{M}[1]{>{\centering\arraybackslash}m{#1}}

\title{A feed-forward neural network as a nonlinear dynamics integrator for supercontinuum generation}

\author[1,*]{Lauri Salmela}
\author[1,2]{Mathilde Hary}
\author[2]{Mehdi Mabed}
\author[3]{Alessandro Foi}
\author[2]{John M. Dudley}
\author[1]{Go{\"e}ry Genty}

\affil[1]{Photonics Laboratory, Physics Unit, Tampere University, 33014 Tampere, Finland}
\affil[2]{Institut FEMTO-ST, Universit\'{e} Bourgogne Franche-Comt\'{e} CNRS UMR 6174, 25000 Besan\c{c}on, France}
\affil[3]{Laboratory of Signal Processing, Tampere University, 33014 Tampere, Finland}

\affil[*]{Corresponding author: lauri.salmela@tuni.fi}

\begin{document}
\twocolumn[
\begin{@twocolumnfalse}
\maketitle
\begin{abstract}
The nonlinear propagation of ultrashort pulses in optical fiber depends sensitively on both input pulse and fiber parameters. As a result, optimizing propagation for specific applications generally requires time-consuming simulations based on sequential integration of the generalized nonlinear Schr{\"o}dinger equation (GNLSE). Here, we train a feed-forward neural network to learn the differential propagation dynamics of the GNLSE, allowing emulation of direct numerical integration of fiber propagation, and particularly the highly complex case of supercontinuum generation. Comparison with a recurrent neural network shows that the feed-forward approach yields faster training and computation, and reduced memory requirements. The approach is generic and can be extended to other physical systems.
\end{abstract}
\end{@twocolumnfalse}
]

Neural networks (NNs) are a central subset of machine learning techniques widely used in data analysis, classification and prediction \cite{goodfellow2016deep}. A central aspect of NNs is their ability to link the input and output of a multidimensional system, of particular benefit for modeling complex and analytically intractable relationships as is typically the case with nonlinear systems. Indeed, several studies have demonstrated the use of NNs to forecast nonlinear evolution using including physics-informed methods  \cite{brunton2016discovering, raissi2018deep, raissi2019physics}, data-driven approaches \cite{vlachas2020backpropagation, jiang2019model, sangiorgio2020robustness}, and hybrid techniques \cite{wikner2020combining, lei2020hybrid}. In optics, NNs are becoming increasingly applied to study multidimensional ultrafast and chaotic systems \cite{genty2020machine}, with recent applications including the optimization of mode-locked lasers \cite{pu2019intelligent, andral2015fiber, kokhanovskiy2019machine} and the analysis of ultrafast instabilities \cite{narhi2018machine, salmela2021predicting, amil2019machine}. 

A particular focus has been the use of NNs to study nonlinear propagation and supercontinuum (SC) generation in optical fibre \cite{narhi2018machine,salmela2020machine,salmela2021predicting}, a complex process involving multiple nonlinear and dispersive effects \cite{dudley2006supercontinuum}. Both the propagation dynamics and the output spectral and temporal characteristics depend sensitively on the injected pulse and fiber parameters, and matching input conditions to achieve a desired output is a complex multivariate problem. The traditional approach for optimization is based on parameter scanning using step-by-step integration of the generalized nonlinear Schr{\"o}dinger equation (GNLSE) \cite{agrawal}. Yet whilst the GNLSE has been shown to accurately model fiber nonlinear dynamics, direct simulations are time consuming, especially with a large parameter space of potential boundary conditions.

To overcome this limitation, attempts have been made to use machine learning techniques to optimize and control fiber dynamics, with one approach being the use of genetic algorithms to tailor broadband SC spectra \cite{wetzel2018customizing, michaeli2018genetic}. More recently, recurrent neural networks (RNNs) using only the input temporal (or spectral) intensity profile of the injected pulse have been shown to emulate fiber propagation dynamics \cite{salmela2021predicting} with accurate prediction of SC evolution maps in computation times as short as one second. A limitation, however, is that this approach requires an initial training phase of several hours due to the multiple iterative loops associated with the RNN internal memory. 

Here, we show how the full-field (intensity and phase) evolution of ultrashort pulses in optical fiber can be accurately modeled with a faster and simpler feed-forward neural network (FNN) over a wide range of input pulse properties (peak power, duration, chirp), and fiber parameters (dispersion, nonlinearity). The key conceptual novelty is that we train the network to learn the  differential propagation dynamics of the GNLSE i.e. to accurately replicate the change in intensity and phase of the electric field between elementary steps. Once trained on the differential  dynamics, the network then can model the long-term evolution from a given input.  We also perform a detailed comparison with a RNN model, highlighting the benefits of the FNN approach in terms of speed and memory. 

\begin{figure}[!t]
  \centering
  \includegraphics[width=\linewidth]{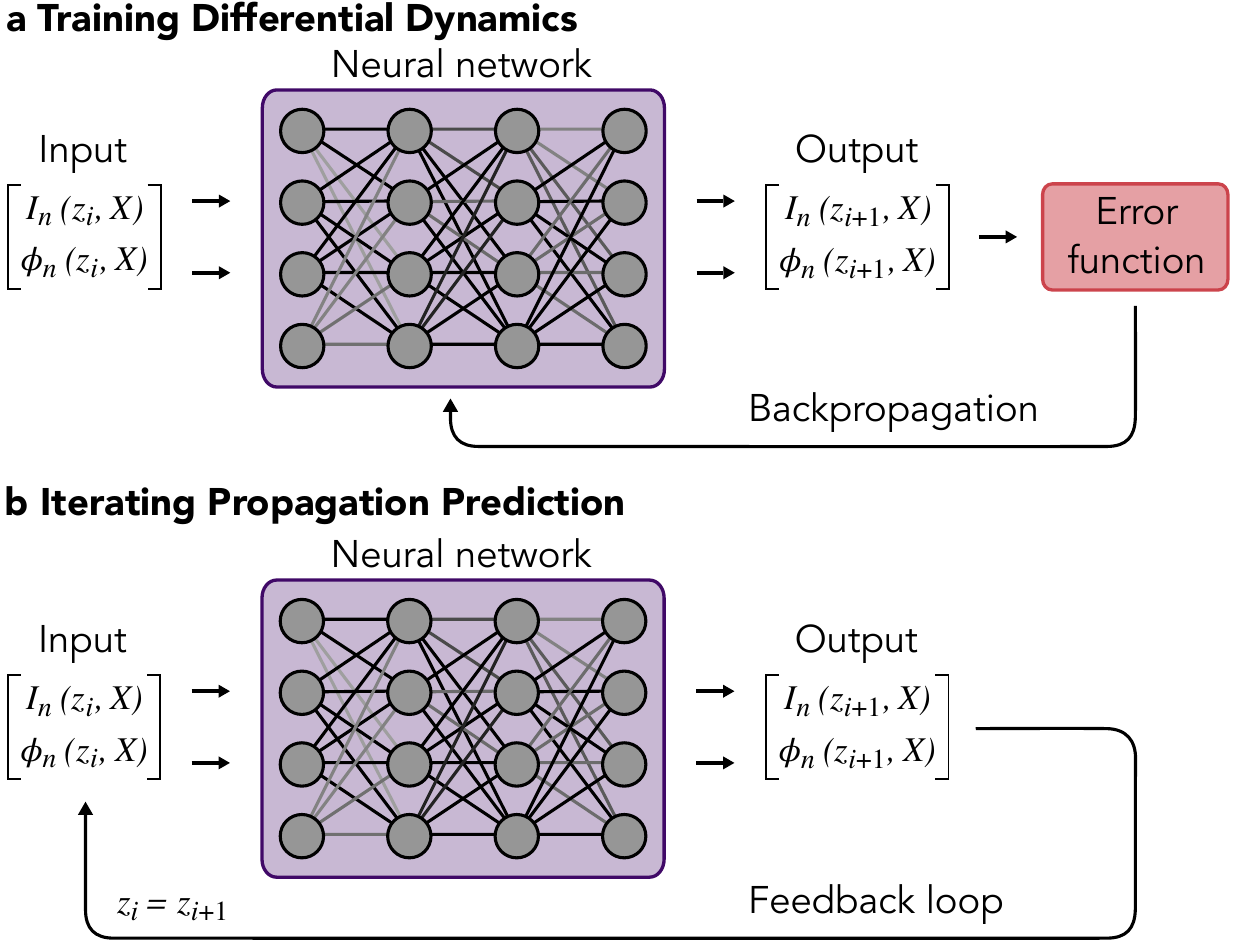}
\caption{Feed-forward neural nonlinear dynamics integrator principle. {\bf a} Training Differential Dynamics. Training is done from multiple input/output pairs generated by direct integration of the GNLSE and corresponding to the temporal ($X=T$) or spectral ($X=\omega$) intensity $I$ and phase $\Phi$ of the propagating field at distances separated by an elementary step $\Delta z$ (see text for details). The network variables are adjusted via gradient descent backpropagation. {\bf b} Iterating Propagation Prediction. Once trained, the network effectively acts a GNLSE integrator and predict iteratively the intensity and phase evolution via feedback loop. The prediction is initialized from the intensity and phase profile at the fiber input.}
\label{fig:schematic}
\end{figure}

The principle is illustrated in Fig. \ref{fig:schematic}. We first generate an ensemble of pulse propagation data corresponding to highly complex broadband coherent SC generation. The dynamical map is completely characterized by a vector $[I_n(z_i,X), \Phi_n(z_i,X]$, where $I_n$ and $\Phi_n$ represent intensity and phase at some distance $z_i$, expressible either in the temporal ($X=T$) or spectral domains ($X=\omega$), and from which the complex electric field can be reconstructed. The subscript $n=1 \ldots N$ indicates a particular map for a distinct set of input pulse and fiber parameters.  

This data is generated by numerically integrating the GNLSE with the split-step method, seeded by hyperbolic-secant input pulses at $\lambda_0=830$~nm, with peak power and duration (FWHM) in the range $P_0=$~0.77--1.43~kW and $T_{\rm FMHM}=$70--130~fs ($\pm$30\% variation). The fiber dispersion parameters are: $\beta_2 = -5.90 \times 10^{-27}$~s$^2$m$^{-1}$, $\beta_3 =  4.21 \times 10^{-41}$~s$^3$m$^{-1}$, $\beta_4 = -1.25 \times 10^{-55}$~s$^4$m$^{-1}$, and $\beta_5 = -2.45 \times 10^{-70}$~s$^5$m$^{-1}$ (zero-dispersion wavelength at 767 nm), and the nonlinear coefficient is: $\gamma=0.1 \rm W^{-1}m^{-1}$. The fiber length is $L$ = 20 cm.

The key idea shown in Fig.~1(a) is to teach the network the differential change in intensity and phase associated with an elementary propagation distance $\Delta z$. To achieve a performance advantage relative to direct integration, the aim is to use a significantly larger step in the FNN compared to that used in GNLSE integration. To this end, the intensity and phase evolution are downsampled at distances $z_i=(i-1)\Delta z$ ($i=1..M$), where \hbox{$\Delta z=L/M$ = 0.1~cm} is 50 times larger than in the GNLSE simulations used to generate the data. The downsampled vectors are then used as the FNN input. The network output vectors after an elementary step $\Delta z$ are $[I_n(z_{i+1},X), \Phi_n(z_{i+1},X)]$. The change in the intensity and phase modeled by the FNN is then compared to that from the GNLSE via an error function \cite{narhi2018machine}.

Once trained, the neural network acts as a very fast and memory-efficient GNLSE integrator. It can predict the intensity and phase $[I(z+\Delta z,X), \Phi(z+\Delta z,X)]$ after an elementary propagation distance $\Delta z$ given the complex field $[I(z,X), \Phi(z,X)]$ at distance $z$, from which the dynamical evolution of the complex electric field can be reconstructed. The trained FNN can then be used to predict propagation dynamics over an extended distance using a iterative loop (see Fig.~1(b)) such that the intensity and phase $[I(z_{i+1},X), \Phi(z_{i+1},X)]$ are fed back to the network as a new input to predict the field amplitude $[I(z_{i+2},X), \Phi(z_{i+2},X)]$ at distance $z+2\Delta z$. This operation is performed over the full propagation distance.

The neural network itself consists of 3 hidden layers of 2000 nodes with ReLU activation $\left(f(x)=\text{max}(0,x)\right)$ and a sigmoid output layer with 2048 nodes. The codes were written in Python using Keras with Tensorflow backend \cite{abadi2016tensorflow}. The network is trained for 80 epochs with RMSprop optimizer and adaptive learning rate. The network can be trained in the temporal or spectral domain and with data input on either linear or logarithmic (dB) scales. In the results shown below we used ensembles of spectral evolution maps in logarithmic scale. Examples of time domain evolution using linear input are in the Supplementary information (Fig.~S1). The accuracy of the network is tested with a separate set of propagation maps not used in the training phase. We quantify performance using the average (normalized) root mean squared (RMS) error:
\begin{equation}
    \label{eq:error}
    R = \sqrt{\frac{\sum_{d,i} (x_{n,d,i} - \hat{x}_{n,d,i})^2}{\sum_{d,i} (x_{n,d,i})^2}},
\end{equation}
where $\mathbf{x}_n$ and $\hat{\mathbf{x}}_n$ denote GNLSE simulation and FNN prediction for a particular realization $n$. Variables $d$ and $i$ indicate summation over intensity (spectral or temporal) and propagation steps, respectively. When evaluating performance over an ensemble, the error is calculated over $N$ distinct evolution maps.

\begin{figure*}[!t]
  \centering
  \includegraphics[width=\linewidth]{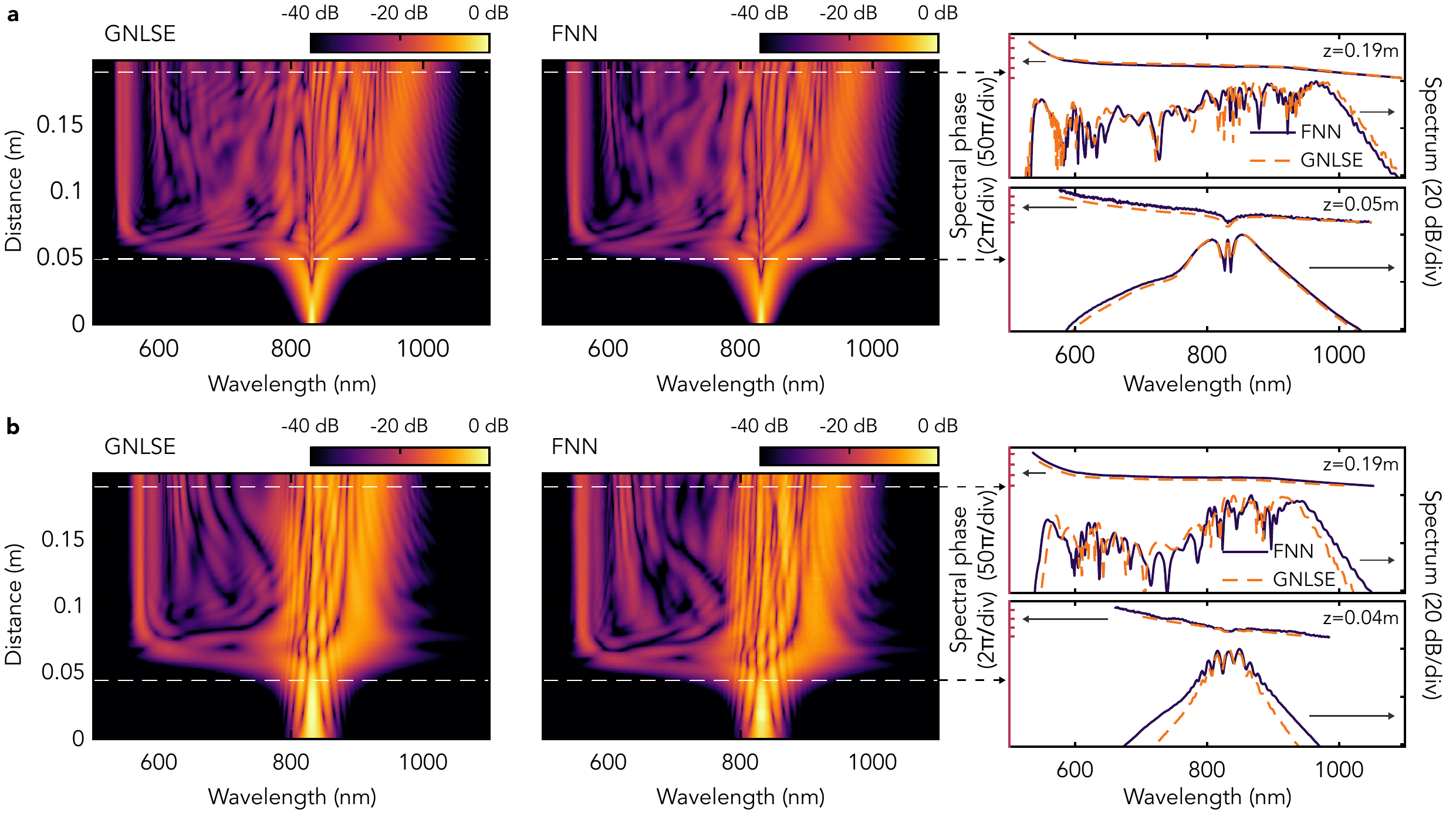}
\caption{Comparison of supercontinuum spectral intensity evolution between GNLSE simulations (left panel) and neural network prediction (FNN, middle panel). The right panel shows the spectral intensity and phase at selected distances as indicated by the arrows. {\bf a} shows the spectral intensity evolution of transform limited input pulses for input peak power and pulse duration values of $P_0=1.32 \rm kW$ and $T_{FWHM}=120 \rm fs$), respectively. {\bf b} shows the spectral intensity evolution for a chirped input pulse for input peak power and pulse duration values of $P_0=942 \rm W$ and $T_{FWHM}=84 \rm fs$) with initial positive chirp of 1.53 times the TL bandwidth.}
\label{fig:SC}
\end{figure*}

We first show the ability of the FNN to predict SC evolution developing from transform-limited input pulses. Here, we used an ensemble of 1500 simulations: 1400 for training phase and 100 for  testing. Predicted spectral evolution maps are shown in Fig.~\ref{fig:SC} for input peak power and pulse duration of 1.32~kW and 120~fs (Fig.~\ref{fig:SC}a). For comparison, we also plot the evolution from direct GNLSE integration. The RMS error for the realizations shown in Fig.~\ref{fig:SC}a is \hbox{R = 0.098}, while the average error computed over the 100 test evolution maps is \hbox{R = 0.094}. The neural network accurately predicts the SC development, with dispersive wave and soliton dynamics reproduced over a $\sim 40$~dB dynamic range.  

We next tested the ability of the network to model SC development from chirped pulses. Here, we performed 3000 simulations with the same parameters as above, except with peak power variation of $\pm$20\% and input pulse spectral bandwidth varying from transform-limited (TL) to twice the TL with random sign of chirp. The SC spectral evolution predicted by the network for pulses with 942~W peak power, 84~fs duration, and positive chirp of 1.53 times the TL bandwidth are shown Fig.~\ref{fig:SC}b. Again in this case we see how the main features including the  spectral interference fine structure are well-reproduced (\hbox{R = 0.190}) by the FNN model although we do note a small discrepancy in the distance of maximum compression at -20~dB bandwidth. The RMS error \hbox{R = 0.383} ({0.242} median) computed over the 100 test ensemble shows that the network also accurately models chirped pulse dynamics. 

\begin{figure}[!t]
  \centering
  \includegraphics[width=\linewidth]{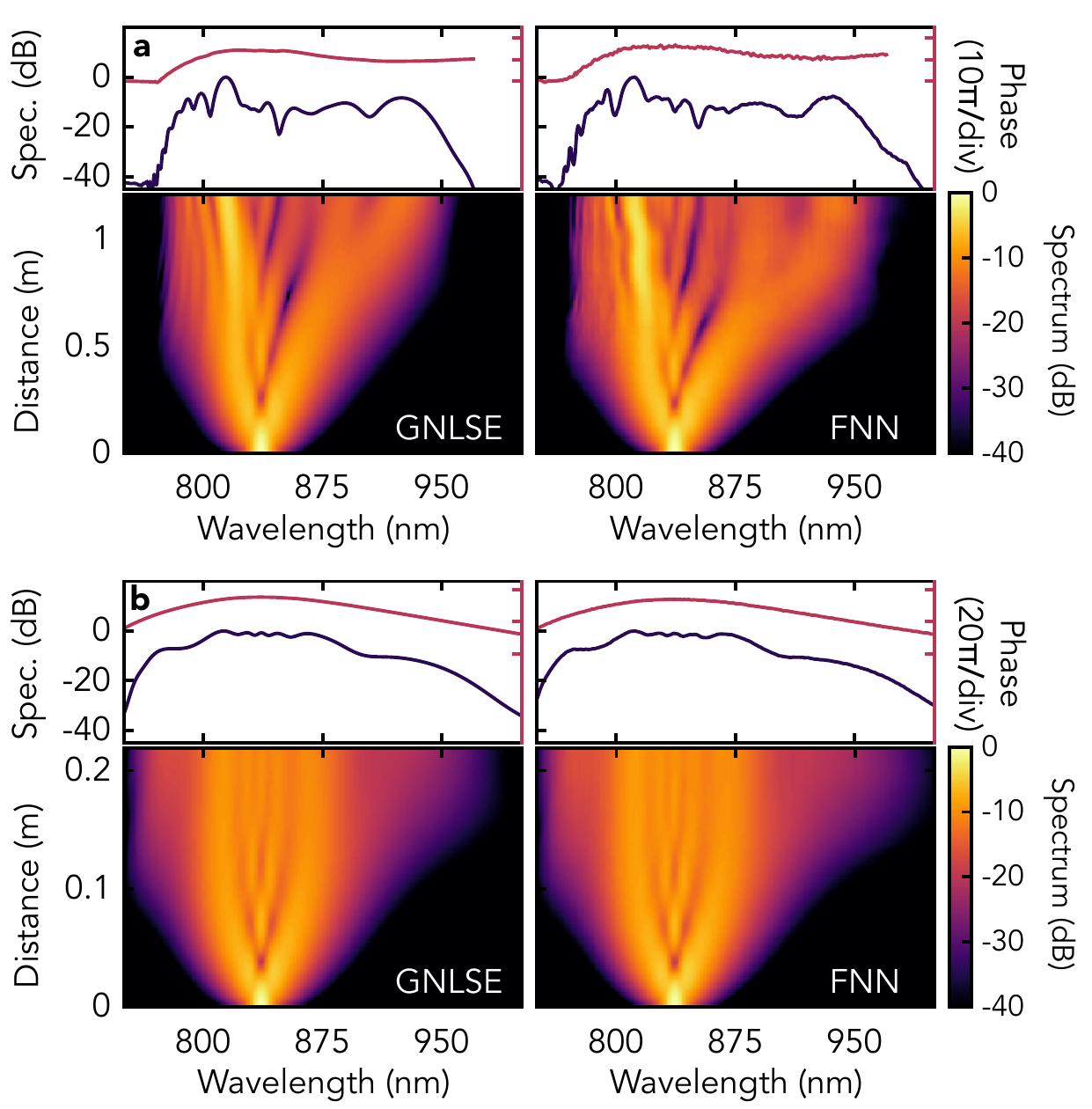}
  \vspace*{-0.5cm}
\caption{Spectral intensity evolution from simulations (GNLSE, left panel) and predicted by the neural network (FNN, right panel) for normal near-zero-dispersion pumping in {\bf a} ($\gamma=0.01~\rm W^{-1}m^{-1}$, $\beta_2=1.3\times10^{-27}~\rm s^2m^{-1}$, $\beta_3=2\times10^{-41} \rm s^3m^{-1}$, $P_0=2.0~\rm kW$, $\lambda_0=835~\rm nm$, $T_{\rm FWHM}=100~\rm fs$) and far-normal pumping in {\bf b} ($\gamma=0.01~\rm W^{-1}m^{-1}$, $\beta_2=7.2\times10^{-27}~\rm s^2m^{-1}$, $\beta_3=2\times10^{-41}~\rm s^3m^{-1}$, $P_0=13.6~\rm kW$, $\lambda_0=835~\rm nm$, $T_{\rm FWHM}=100~\rm fs$). The top panels show the spectral intensity and phase at the fiber output.}
\label{fig:normal}
\end{figure}

The results above correspond to the case of anomalous dispersion regime SC generation, but the network can be trained over a much wider range of dynamics. To this end, one can use the normalized form of the GNLSE to generate the training ensemble of evolution maps (see Supplementary information), and map dimensional parameters to normalized values to predict the evolution corresponding to a specific set of parameters. For example, Fig.~\ref{fig:normal}  plots examples of predicted SC evolution for a pump wavelength in the normal dispersion region (see caption for parameters). Specifically, Fig.~\ref{fig:normal}a shows results for a TL limited pulse injected near the zero-dispersion wavelength while Fig.~\ref{fig:normal}b shows the spectral evolution for a pump wavelength further detuned into the normal dispersion regime. We observe very good accuracy with \hbox{R = 0.141} for Fig.~\ref{fig:normal}a and \hbox{R = 0.043} for Fig.~\ref{fig:normal}b. The RMS error over an ensemble of 200 realizations is \hbox{R = 0.060}. Predictions in the time domain can be found in the Supplementary information (Fig. S2).

To reduce computational memory and increase the speed in the training phase, one can train the network from convolved spectral intensity and phase evolution maps. At first sight, a disadvantage of using convolved data is that the resulting wavelength/frequency grid is no longer on a Fourier grid, requiring separate training to predict spectral and temporal evolution. However, this is in fact a major benefit, because it allows to appropriately select the resolution in the spectral or temporal domains to optimally capture the relevant physical structure. 

\begin{figure}[!t]
  \centering
  \includegraphics[width=\linewidth]{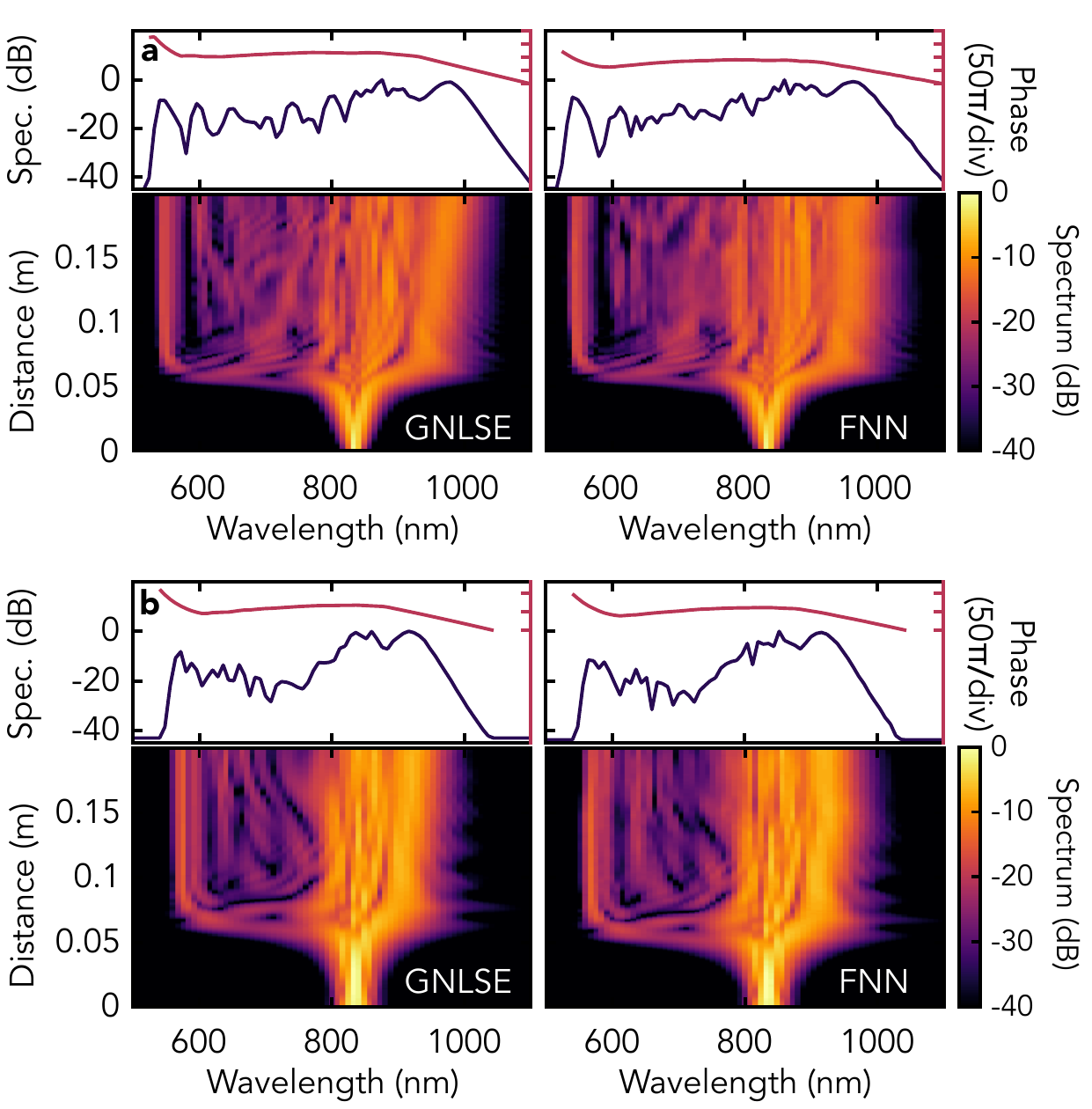}
  \vspace*{-0.5cm}
\caption{Comparison between the spectral evolution from simulations (GNLSE, left panel) and predicted by the neural network (FNN, right panel) when using convolved evolution maps for training with parameters identical to those in Fig.~\ref{fig:SC}a and b. The top panels show the spectral intensity and phase at the fiber output.}
\label{fig:convo}
\end{figure}

Results of predicted spectral evolution maps using convolved spectral intensity and phase training data with a 8~nm FWHM super-Gaussian spectral filter are shown in Fig.~\ref{fig:convo}a,b. These results correspond to the same input pulse and fiber parameters as in Fig.~\ref{fig:SC}a,b.
We see how the network predictions remain accurate with a mean convolved (logarithmic) spectral intensity RMS of 0.06 and 0.16 calculated over 100 distinct test evolution maps for the transform-limited and chirped cases, respectively. Other examples of predictions using different spectral resolution can be found in the Supplementary information Fig.~S3. 

We then conducted a full comparison test of the computation resources and performance between the FNN model and an RNN similar to that used in Ref. \cite{salmela2021predicting}. The comparison was performed over an ensemble of 12,000 (11,800 for training and 200 for testing) convolved SC evolution maps in the anomalous dispersion regime with variations in peak power, pulse duration and dispersion (see Supplementary information).  Table~\ref{table:comparison} summarizes the results, with examples of predicted maps shown in the Supplementary information (Fig.~S4). For completeness we also list the computational resources used by the GNLSE simulations. Both FNN and RNN used the same number of free parameters/network variables, but the RNN is trained from spectral intensity maps which reduces by half the number of grid points compared to the FNN that includes both intensity and phase. The computational advantage of the FNN is clear. Specifically, FNN training and simulation times are reduced by a factor of four and five respectively, while memory usage during training is decreased by a factor of two. As might be expected from the faster computation, the FNN does show increased error compared to the RNN, but this does not lead to any significant visual differences in the evolution maps seen in the figures.

\begin{table}[!t]
\centering
\begin{tabular}{m{2.2cm} M{1.5cm} M{1.5cm} M{1.5cm}}
 & GNLSE & RNN & FNN \\
 \hline
 RMS error &  N/A & R = 0.09  & R = 0.19 \\
 \hline
 Training time* & N/A &  7.7 h  & 1.9 h \\
 \hline
 Simulation time** & 38 min & 1.6 s & 0.35 s \\
 \hline
 Memory* & 79 GB & 7.7 GB & 3.2 GB \\
 \hline
 Network var. & N/A & 600k & 600k \\
 \hline
 Num. points &  8,192 & 132 & 264 \\
 \hline
 \end{tabular}
 \vspace*{-0.1cm}
\caption{Comparison between normalized GNLSE numerical simulations, recurrent neural network (RNN) \cite{salmela2021predicting}, and feed-forward neural network (RNN) for convolved spectral data.
*11800 sims **200 sims}
\label{table:comparison}
\end{table}

These results have shown model-free prediction of the full-field dynamics of ultrashort pulse propagation in optical fiber based on a feed-forward neural network trained to recognize differential propagation dynamics within a GNLSE model. As compared to the recently introduced RNN approach, this FNN method is simpler and possesses significant advantages in terms of speed and memory. We expect our results to be of significance for the real-time optimization and control of nonlinear dynamics and we anticipate this approach could become a standard tool in nonlinear physics.
\medskip

\noindent\textbf{Funding.} Faculty of Engineering and Natural Sciences graduate school of Tampere University. French Agence Nationale de la Recherche (ANR-15-IDEX- 0003, ANR-17-EURE-0002). Academy of Finland (298463, 318082, Flagship PREIN 320165).

\noindent\textbf{Disclosures.} The authors declare no conflicts of interest.

\noindent\textbf{Data availability.} The data in this paper may be obtained from the authors upon request.

\noindent See Supplement 1 for supporting content.

\bibliographystyle{ieeetr}
\bibliography{refs}

\clearpage

\setcounter{figure}{0}
\makeatletter 
\renewcommand{\thefigure}{S\@arabic\c@figure}
\makeatother

\onecolumn
\noindent\textbf{\LARGE Supplementary information}

\section*{Supplementary results}

\noindent\textbf{Same evolution scenarios as in Figure 2 of the main manuscript but with training and prediction performed in the time domain.} The RMS errors for temporal intensity are \hbox{R = 0.54} and \hbox{R = 0.70} for the transform-limited and chirped supercontinuum cases, respectively.

\begin{figure}[!ht]
  \centering
  \includegraphics[width=\linewidth]{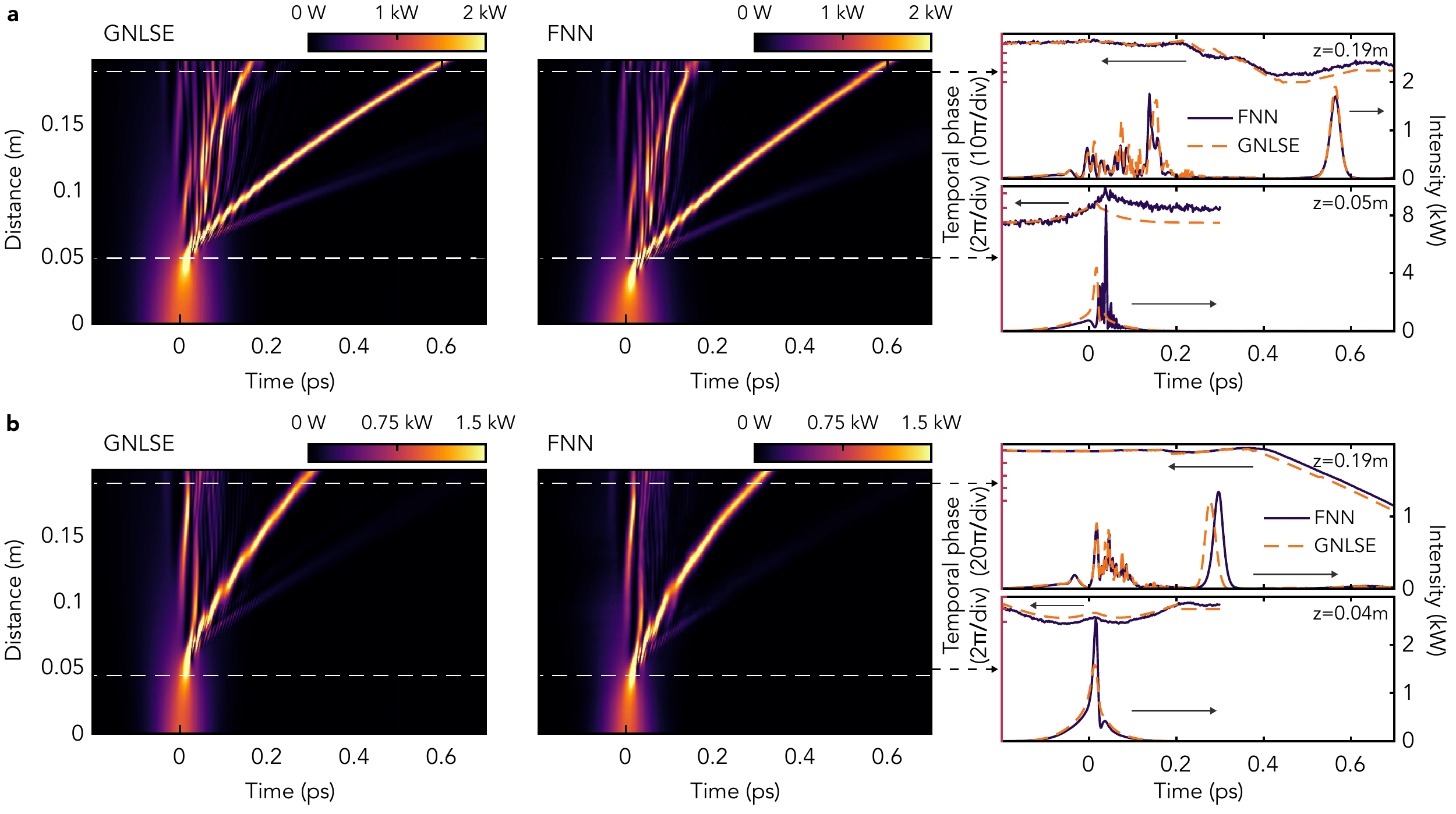}
  \caption{Temporal intensity evolution of supercontinuum from simulations (GNLSE, left panel), predicted by the neural network (FNN, middle panel), and their comparison at selected distances (right panel). {\bf a} shows the evolution of a transform limited input pulse, and  {\bf b} shows the evolution for a chirped input pulse (see main text for details).}
\label{fig:SCtime}
\end{figure}

\noindent\textbf{Same evolution scenarios as in Figure 3 of the main manuscript but with training and prediction performed in the time domain}  The RMS error is \hbox{R = 0.17} over 200 test evolution maps.

\begin{figure*}[!ht]
  \centering
  \includegraphics[width=\textwidth]{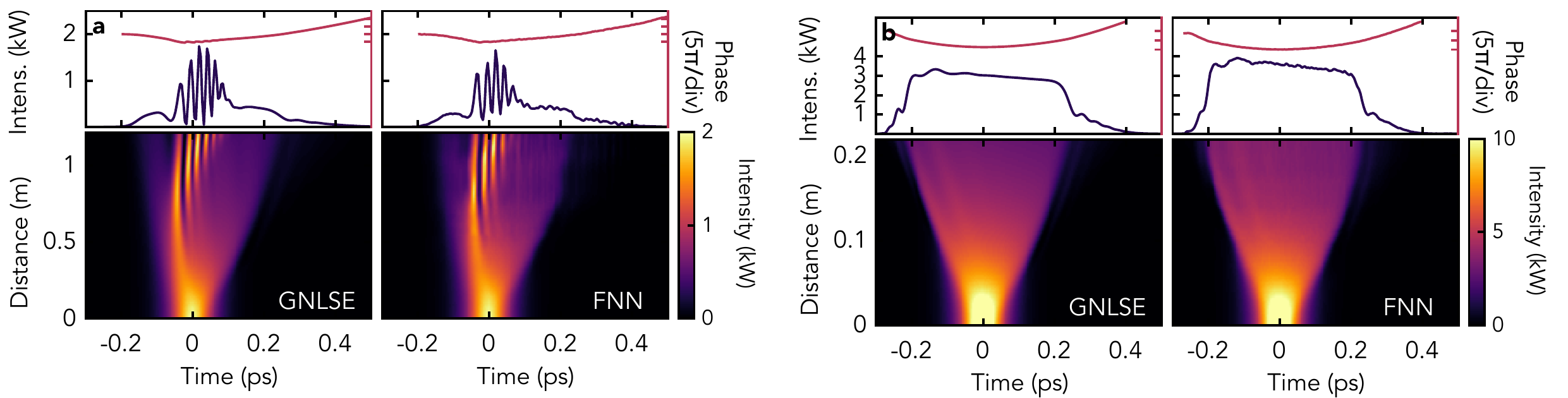}
\caption{Temporal intensity evolution of supercontinuum from simulations (GNLSE, left panel), predicted by the neural network (FNN, right panel) for normal near-zero-dispersion pumping in {\bf a} and far-normal pumping in {\bf b}. The top panels show the temporal intensity and phase at the fiber output}
\label{fig:normaltime}
\end{figure*}

\noindent\textbf{Same evolution scenarios as in Figure 4a,b but with training intensity and phase data convolved with a 4~nm FWHM super-Gaussian filter.}
The (logarithmic) RMS error with the 4~nm resolution is \hbox{R = 0.06} and \hbox{R = 0.17} for the transform-limited and chirped cases, respectively, similar values as observed with the 8~nm resolution.

\begin{figure*}[!ht]
  \centering
  \includegraphics[width=\textwidth]{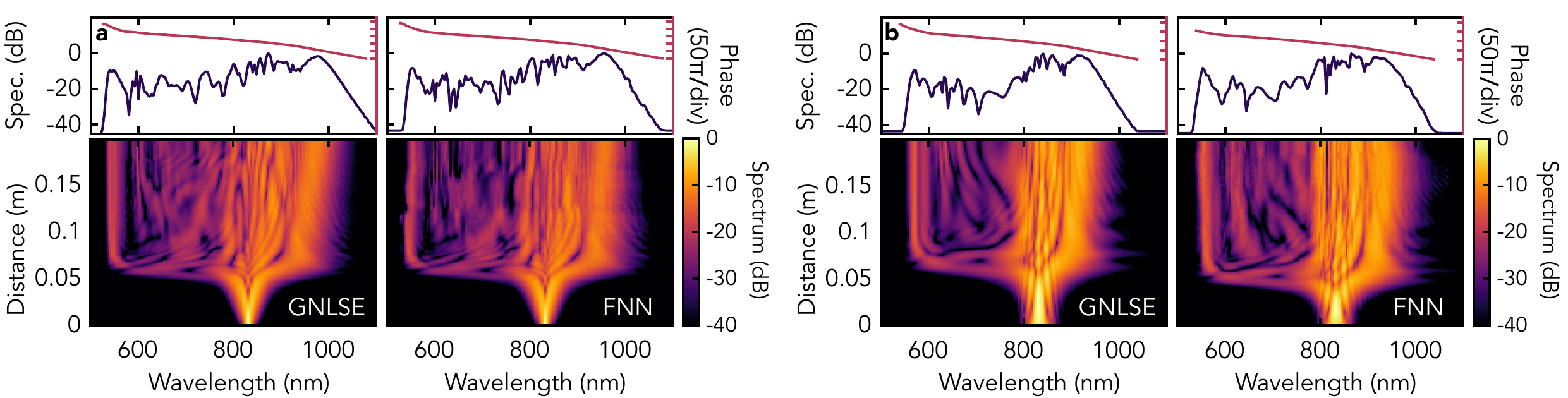}
\caption{Convolved spectral intensity evolution of supercontinuum from simulations (GNLSE, left panel), predicted by the neural network (FNN, right panel) and corresponding to the cases illustrated in Fig.~4a and b. The top panels show the spectral intensity and phase at the fiber output.}
\label{fig:convo4nm}
\end{figure*}

\noindent\textbf{Comparison between recurrent and feed-forward neural network} The (logarithmic) RMS errors over 200 test realizations are \hbox{R = 0.09} and \hbox{R = 0.19} for the recurrent and feed-forward neural networks, respectively.

\begin{figure*}[!ht]
  \centering
  \includegraphics[width=\textwidth]{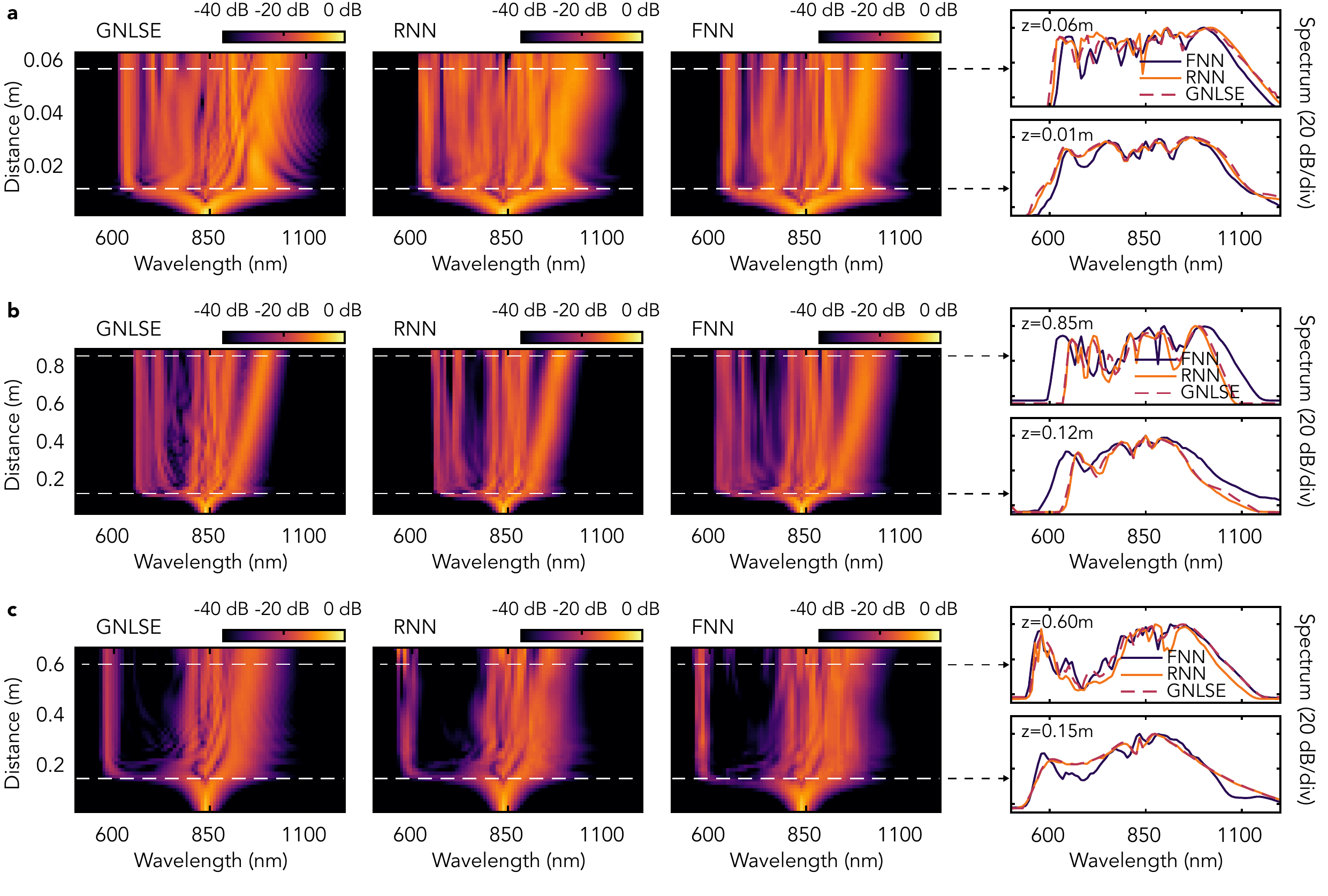}
\caption{Convolved spectral intensity evolution of supercontinuum from simulations (GNLSE), predicted by the recurrent (RNN) and neural network (FNN), and their comparison at selected distances (right panel). {\bf a} shows the predictions for sech-type input pulse centered at 830~nm with 7.6~kW peak power and 40~fs duration with fibre parameters of $\gamma = 0.1$~W$^{-1}$m$^{-1}$, $\beta_2 = -8 \times 10^{-27}$~s$^2$m$^{-1}$ and $\beta_3 = 9 \times 10^{-41}$~s$^3$m$^{-1}$. {\bf b} shows the results for an input pulse with 2.9~kW peak power and 120~fs duration, and $\gamma = 0.0184$~W$^{-1}$m$^{-1}$, $\beta_2 = -5.1 \times 10^{-27}$~s$^2$m$^{-1}$ and $\beta_3 = 4.3 \times 10^{-41}$~s$^3$m$^{-1}$. {\bf c} shows the results for a realization with 3.0~kW peak power and 60~fs duration, and $\gamma = 0.01$~W$^{-1}$m$^{-1}$, $\beta_2 = -1.7 \times 10^{-27}$~s$^2$m$^{-1}$ and $\beta_3 = 6.5 \times 10^{-42}$~s$^3$m$^{-1}$.}
\label{fig:comparison}
\end{figure*}

\section*{Numerical simulations}

\noindent\textbf{GNLSE simulation.} 
 We model supercontinuum generation by injecting sech-type transform-limited pulses with 0.77--1.43~kW peak power and 70--130~fs duration (FWHM) ($\pm$30\% variation) at 830~nm center wavelength are injected into the anomalous dispersion regime of a 20~cm nonlinear fibre, including higher-order dispersion, self-steepening and Raman effect.
The nonlinear coefficient of the fibre is $\gamma = 0.1$~W$^{-1}$m$^{-1}$, and the Taylor-series expansion coefficients for dispersion at 830~nm are 
$\beta_2 = -5.90 \times 10^{-27}$~s$^2$m$^{-1}$,
$\beta_3 =  4.21 \times 10^{-41}$~s$^3$m$^{-1}$,
$\beta_4 = -1.25 \times 10^{-55}$~s$^4$m$^{-1}$, and
$\beta_5 = -2.45 \times 10^{-70}$~s$^5$m$^{-1}$.
The simulations use 1024 spectral/temporal grid points with temporal window size of 2~ps, and a step size of 0.02~mm (10,000 steps).
For the neural network, the propagation is downsampled at a constant propagation step of \hbox{$\Delta z$ = 0.1~cm}, yielding 200 propagation steps. Shot noise is added via one-photon-per-mode with random phase in the frequency domain, although
noise does play no significant physical role in the regime of coherent propagation studied here.

\noindent\textbf{Normalized GNLSE.} 
For generalization of the the prediction model, we used the normalized form of the GNLSE:
\begin{equation*}
    \label{eq:GNLSE}
    i\frac{\partial\psi}{\partial\xi}
    - \frac{\text{sgn}(\beta_2)}{2}\frac{\partial^2\psi}{\partial\tau^2}
    - i\frac{q}{6}\frac{\partial^3\psi}{\partial\tau^3} 
    + \left(1+ i s \frac{\partial}{\partial\tau} \right)
    \left(\psi \int_{-\infty}^{+\infty} r(\tau\prime) |\psi(\tau-\tau\prime) \xi|^2 \text{d}\tau\prime \right)
    = 0,
\end{equation*}
where $\tau = T/T_0$, $\xi = zT_0^2/|\beta_2|$ and $\psi (\xi,\tau)= NA(z,T)/\sqrt{P_0}$,  $q=\beta_3/|\beta_2|T_0$, $s=1/\omega_0T_0$, and $r$ are the normalized time, propagation distance, amplitude, third-order dispersion,  shock-term, and Raman response, respectively. 

For the normal dispersion, we model the propagation of transform-limited hyperbolic-secant pulses centered at 835~nm in the normally dispersive fiber (i.e. \hbox{$sgn(\beta_2) = 1$)}. Simulations included a variation to the soliton number from 2 to 8 and dispersion parameter from 1 to 25 with the positive sign of $\beta_2$. We used 512 temporal grid points with temporal window size of 400 (normalized units) and normalized distance set to 0.5 with 9,000 steps.

For the comparison with the recurrent neural network \cite{salmela2021predicting}, we simulated the propagation of transform-limited hyperbolic-secant pulses centered at 830~nm in the anomalous dispersion regime (i.e. $sgn(\beta_2) = -1$). The soliton number, pulse duration and third-order dispersion parameter were randomly varied in the interval 2 to 8, 30 to 130~fs and 1 to 9, respectively. The simulations used 8192 temporal grid points with temporal window size of 350 (normalized units) and normalized distance set to 2 with 9,000 steps.

\end{document}